\documentclass[prl,twocolumn,showpacs,showkeys,amsmath,amssymb,floatfix]{revtex4}
\usepackage{graphicx}
\usepackage{dcolumn}
\usepackage{eucal}
\usepackage{amsfonts}
\usepackage{latexsym}
\usepackage{epsfig}
\usepackage{color}

\usepackage{amsfonts,amssymb,eucal}
\usepackage{amsthm}
 \usepackage{amsmath}
\usepackage{amscd}
 \usepackage{latexsym} 
 
\begin{document}
\title{A new method for  a global fit of the CKM matrix}
\author{Petre Di\c t\u a$^{1}$\footnote {dita@zeus.theory.nipne.ro}}
\affiliation{$^{1}$ Institute of Physics and Nuclear Engineering,
P.O. Box MG6, Bucharest, Romania}
\begin{abstract}\noindent
We report on a new method to a global fit of the CKM matrix by using the necessary and sufficient condition the data have to satisfy in order to find a unitary matrix compatible with them. This condition writes as $-1\le \cos\varphi\le 1$ where $\varphi$ is the phase that accounts for  CP violation. By using it we get that the experimental data are to a high degree compatible to unitarity and that $\varphi$ takes values around $90^0$, in contrast  to the previous determinations.
Numerical results are provided for the CKM matrix entries, the mixing angles between generations and all the angles of the standard unitarity triangle.
\end{abstract}
\maketitle
 The determination of the Cabibbo-Kobayashi-Maskawa matrix that parameterizes
the weak charged current interactions of quarks is a lively subject in particle physics. The four independent parameters of this matrix govern all flavor changing transitions of quarks in the Standard Model, and their determination is an
important task for both experimenters and theorists. The large interest in the subject is also reflected by the workshops organized in the last two years whose main subject was the CKM matrix \cite{BBGS, AB, BFKS}.

Although it seems that at the level of experts there exists a consensus concerning  the method of  extracting from experimental data information about the CP violating phase \cite{BBGS}-\cite{HLLL1}, the usual method of considering only  the orthogonality of the first and third columns is not reliable as we shall show in the paper.
The main shortcomings of the current approach are the following: a) one uses 
 one unitarity triangle instead of six; b) the currently used four independent parameters are not re-phasing invariant; c) one works with an approximation of the standard parameterization which may lead to inconsistencies. 

 The 
 aim of this paper is to propose an alternative method for imposing unitarity
by   fully exploiting  the constraints implied by it. We consider that the second-generation B-decay experiments of the LHC era, when the accuracy will make a tremendous difference, require to keep improving all the tools we are working with. Our approach provides 
the necessary and sufficient condition the data have to satisfy in order to find a unitary matrix compatible with them; this condition is given by
$-1\leq\cos\varphi\leq 1$, where $\varphi$ is the phase entering the CKM matrix. We construct a theoretical model and we use it to test
 the  unitarity property of the data  as they are provided by the Particle Data Group (PDG) \cite{Ha}. We find that one can reconstruct from PDG data a unitary matrix and the phase $\varphi$  is close to  $\pi/2$.

Our theoretical  framework is as follows.
 We use the standard parameterization advocated by  PDG that we write it in a
completely rephaising invariant form as
\begin{eqnarray}
U_{CKM}=\label{ckm}
\end{eqnarray}
{\footnotesize
\begin{eqnarray}
 \left(\begin{array}{ccc}
c_{12}c_{13}&c_{13}s_{12}&s_{13}\\
-c_{23}s_{12}e^{i \varphi}-c_{12}s_{23}s_{13}&c_{12}c_{23}e^{i \varphi}-s_{12}s_{23}s_{13}&s_{23}c_{13}\\
s_{12}s_{23}e^{i\varphi}-c_{12}c_{23}s_{13}&-c_{12}s_{23}e^{i \varphi}-s_{12}c_{23}s_{13}&c_{23}c_{13}
\end{array}\right)\nonumber \\\nonumber
\end{eqnarray}
}with $c_{ij}=\cos \theta_{ij}$ and  $s_{ij}=\sin \theta_{ij}$
for the generation labels $ij=12, 13, 23$, and $\varphi$ is the phase that encodes the breaking of the $CP$-invariance. 
It is easily seen that by multiplying a unitary matrix at left and at right with arbitrary diagonal phase matrices, the phase invariance implies that we can freely choose  the entries of one column and of  one row as non-negative numbers. We used this property for rewriting the standard parameterization \cite{CK} in the above form. We remind that  phase invariance of the $U_{CKM}$-transformed quark wave functions is a requirement for physically meaningful quantities.

On the other hand from experiments one measures a matrix whose entries are positive
\begin{eqnarray}
V=\left(\begin{array}{ccc}
\vspace*{1mm}
V_{ud}^2&V_{us}^2&V_{ub}^2\\
\vspace*{1mm}
V_{cd}^2&V_{cs}^2&V_{cb}^2\\

V_{td}^2&V_{ts}^2&V_{tb}^2\\
\end{array}\right)\label{pos}
\end{eqnarray}
entries that, in principle, can be determined from the weak decays of the relevant quarks, and/or from deep inelastic neutrino scattering \cite{Ha}. In other words we make a clear distinction between the unitary CKM matrix $U_{CKM}$ and the {\em positive} matrix $V$ provided by the data. 
The main theoretical problem is to see if from a matrix as (\ref{pos}) one can reconstruct a unitary matrix as (\ref{ckm}). If the experimental data are compatible with unitarity  the weakest form of this property is expressed as  follows

\begin{eqnarray}
\sum_{i=d,s,b} V_{ji}^2-1=0, \quad j=u,c,t\nonumber \\
\sum_{i=u,c,t} V_{ij}^2-1=0, \quad j=d,s,b \label{sto}
\end{eqnarray}

We stress that the above relations does not test the unitarity, as it is 
usually stated in many papers; they are  necessary but not sufficient conditions. The class of positive matrices satisfying  Eqs. (\ref{sto}) is considerable larger than the class of positive matrices coming from unitary matrices. The set  (\ref{sto}) is known in the mathematical literature as doubly stochastic matrices, and the subset coming from  unitary matrices $V_{ij}^2=|U_{ij}|^2$ is known as unistochastic ones \cite{MO}. The double stochastic matrices have an important property, they are a convex set, i.e. if $V_1$ and $V_2$ are doubly stochastic so  is their convex combination $\alpha\,V_1+(1-\alpha)V_2$, $\alpha \in [0,1]$ as it is easily checked.

The first problem to solve is to find a necessary and sufficient criterion for discrimination between the two sets. Assuming  the experimental data (\ref{pos}) come from a unitary matrix as (\ref{ckm}) we obtain the following relations between the parameters entering $U_{CKM}$ and $V$ matrices

\begin{eqnarray}\begin{array}{l}
V_{ud}^2=c^2_{12} c^2_{13},\,\, V_{us}^2=s^2_{12}c^2_{13},\,\, V_{ub}^2=s^2_{13},\\
\\
V_{cb}^2=s^2_{23} c^2_{13},\,\,
 V_{tb}^2=c^2_{13} c^2_{23},\\
\\
V_{cd}^2=s^2_{12} c^2_{23}+s^2_{13} s^2_{23} c^2_{12}+2 s_{12}s_{13}s_{23}c_{12}c_{23}\cos\varphi\\
\\
V_{cs}^2=c^2_{12} c^2_{23}+s^2_{12} s^2_{13} s^2_{23}-2 s_{12}s_{13}s_{23}c_{12}c_{23}\cos\varphi \\
\\
V_{td}^2=s^2_{13}c^2_{12}c^2_{23}+s^2_{12}s^2_{23}-2 s_{12}s_{13}s_{23}c_{12}c_{23}\cos\varphi \\
\\
V_{ts}^2=s^2_{12} s^2_{13} c^2_{23}+c^2_{12}s^2_{23} +2 s_{12}s_{13}s_{23}c_{12}c_{23}\cos\varphi \\
\end{array}\label{unitary}
\end{eqnarray}
It is easily seen that the parameterization (\ref{unitary}) satisfies identically the relations (\ref{sto}), and  also   that  $CP$-violation requires
\begin{eqnarray}
\theta_{ij}\neq 0,\,\,ij=12, 13, 23,\,\, \theta_{12}\neq \pi/2,\,\, {\rm and} \,\, \theta_{23}\neq \pi/2\nonumber\\
\end{eqnarray}
We  deduce from  the  factors that multiply $\cos\varphi$ in Eqs.(\ref{unitary}) that an entire region around $0$ and $\pi/2$ is forbidden for the above
parameters. Using the numerical values from PDG data one can get values for $\cos\varphi$ outside
the ``physical'' range $[-1,1]$, or even  complex as we will show in the following. The above relations provides   the necessary and sufficient condition the data have to satisfy in order  the matrix (\ref{pos}) comes from a unitary matrix, and this condition is
\begin{eqnarray}
 -1\leq\cos\varphi\leq 1\label{unit1}
\end{eqnarray}
Since in relations (\ref{unitary}) $\varphi$ enters only in the cosine function we can take $\varphi\in[0,\pi]$ without loss of generality.

The last four relations (\ref{unitary}) provide us formulas for $\cos\varphi$ and these formulas have to give the same number when comparing theory with experiment, by supposing the data come from a unitary matrix. Their explicit form depends on the independent four parameters  we choose to parameterize the data, and we will always choose these parameters as four experimentally measurable quantities, i.e. $V_{ij}^2$. An other reason is that  $V_{ij}^2=|U_{CKM}^{ij}|^2$ being square of the moduli functions they are rephaising invariant; the CP violating phase $\varphi$ does not share this property \cite{DGW}-\cite{NP}. By consequence the parameters entering on the right hand side  of the  relation
\begin{eqnarray}
s_{13}\,e^{-i\delta}=A\,\lambda^3(\rho-i\,\eta)\nonumber
\end{eqnarray}
usually used by the physics community \cite{BLO,HLLL1} {\em are not} rephaising invariant because  the left hand side  is not.
Another invariant parameter is the Jarlskog invariant $J$ \cite{J}, but it it not a measurable quantity. Depending on the explicit choice of the four independent parameters we get one, two, three or four  different expressions for $\cos\varphi$; e.g. if we take $V_{ud},\,V_{us},\,V_{cd},\,V_{cs},$ as independent parameters we get

\begin{eqnarray}
s_{12}&=&\frac{V_{us}}{\sqrt{V_{ud}^2+V_{us}^2}}\,,\,\, s_{13}=\sqrt{1-V_{cd}^2-V_{cs}^2}\,,\,\,{\rm and}\nonumber \\
s_{23}&=&\frac{\sqrt{1-V_{cd}^2-V_{cs}^2}}{\sqrt{V_{ud}^2+V_{us}^2}}
\\\nonumber\label{mix}
\end{eqnarray}

\noindent
and from the sixth Eqs.(\ref{unitary}) we have
\begin{widetext}
\begin{center}
\begin{eqnarray}
\cos\varphi=
\frac{V_{ud}^2+V_{cd}^2 V_{ud}^2+V_{cs}^2 V_{ud}^2 +V_{ud}^4- V_{cs}^2 V_{ud}^4+ V_{us}^2-
V_{cd}^2 V_{us}^2- V_{cs}^2 V_{us}^2 + V_{cd}^2 V_{ud}^2 V_{us}^2
 -  V_{us}^4   -
 V_{cs}^2 V_{ud}^2 V_{us}^2 +V_{cd}^2 V_{us}^4}
{\,2\, V_{ud}\, V_{us}\, \sqrt{1-V_{cd}^2-V_{cs}^2}\,\,\sqrt{1-V_{ud}^2-V_{us}^2}\,\,
\sqrt{V_{ud}^2+V_{us}^2+V_{cd}^2+V_{cs}^2-1}}
\end{eqnarray}
\end{center}
\end{widetext}
 In this case, other two independent formulas are given by the last two equations in relations (\ref{unitary}), and they have to give (almost) identical numerical results if the data are compatible with unitarity. In general the data will give different numerical values for the three functions expressing  $\cos\varphi$. If the independent parameters are $V_{ud},V_{ub},V_{cd},V_{cb}$, i.e. we use the information contained in the first and the third columns, we obtain four different expressions for  $\cos\varphi$, and in this case the mixing angles $\theta_{ij}$ are given by 

\begin{eqnarray}
s_{12}&=&\frac{\sqrt{1-V_{ud}^2-V_{ub}^2}}{\sqrt{1-V_{ub}^2}},\,\,
s_{13}=V_{ub},\nonumber\\
s_{23}&=&\frac{V_{cb}}{\sqrt{1-V_{ub}^2}}
 \label{mix1}
\end{eqnarray}

Looking at Eqs. (7) and  (\ref{mix1}) we see that the expressions defining the mixing angles are quite different. Thus if the data are compatible to the existence of a unitary matrix these mixing angles have to be equal, and this is  a necessary condition for unitarity. As a warning what we said before can be summarized as follows: the unitarity property is a property {\em of all the CKM matrix elements} and not the property of a row and/or a column, as it is   considered by many people working in the field.

To better understand the above considerations, let us consider the toy model  defined by $V_{ij}^2=1/3$. It is easily seen that no matter how we choose the independent variables we get the same numerical results for $s_{ij}$ when we use the {\em exact values} $V_{ij}^2=1/3$. In this case we have
\[
s_{12}=s_{23}=\frac{1}{\sqrt{2}},\,s_{13}=\frac{1}{\sqrt{3}}
\] 
and from any of the last four equations (\ref{unitary}) we get $\cos\varphi=0$, i.e. $\varphi=\pi/2$.  Then from Eq.(1) we obtain  the unitary matrix
\begin{eqnarray} 
U_{toy}=\frac{1}{\sqrt{3}}\left(
\begin{array}{ccc}
1&1&1\\
e^{2\pi i/3}&e^{4\pi i/3}&1\\
e^{4\pi i/3}&e^{2\pi i/3}&1\\
\end{array}\right)\label{toy}
\end{eqnarray}
which is the  matrix of the finite Fourier transform in $d=3$ dimensions. The above matrix has the property that it maximizes the  Jarlskog invariant $J=s_{12}s_{13}s_{23}c_{12}c_{13}^2c_{23}\sin\,\varphi$, $J_{max}=1/(6\sqrt{3})$.
To see that the orthogonality properties of columns and rows are not trivial, let us suppose that the second column of the $U_{toy}$ matrix has the form$(1,1,1)^t$, where $t$ means transpose. It is easily seen that the  orthogonality of the first column with the second and third ones of this new matrix  are exactly satisfied, but the second and the third columns {\em are not orthogonal} ! The same happens with all the rows, i.e. the orthogonality property is very strong implying supplementary constraints that   until now  were
never been  used at their  maximum potential. In the real case of the experimental data the situation is worse since
we use approximate values for  $V_{ij}$ and
then the values of $s_{ij}$ depend on the choice of the independent parameters that define the model and are no more independent of them  as in the case of the $U_{toy}$ matrix.

Now we define a test function that should take into account the double
stochasticity property expressed by the 
conditions (\ref{sto}) and the fact that in general the numerical values of data are such that
  $\cos\varphi$ depends on the choice of the four independent parameters. Our proposal is
\begin{eqnarray}
\chi^2_1=\sum_{i < j}(\cos\varphi^{(i)} -\cos\varphi^{(j)})^2+\sum_{j=u,c,t}\left(
\sum_{i=d,s,b}V_{ji}^2-1\right)^2\nonumber\\
+\sum_{j=d,s,b}\left(
\sum_{i=u,c,t}V_{ij}^2-1\right)^2\hspace*{3cm}\label{chi} 
\end{eqnarray}
Testing our model (\ref{chi}) on the toy matrix (\ref{toy}) we found that by adding a second component of the form
\begin{eqnarray}
\chi^2_2=\sum_{i=u,c}\,\,\sum_{j=d,s,b}\left(\frac{V_{ij}-\widetilde{V}_{ij}}{\sigma_{ij}}\right)^2\label{chi2} 
\end{eqnarray}
the results improve, 
where $\widetilde{V}_{ij}$ is  a numerical matrix that describes the experimental data, and $\sigma$ is the matrix of errors associated to $\widetilde{V}_{ij}$. The test has also shown that better results are obtained when the number of $\cos\varphi$ is  large.
In the last sum we use only the data coming from the  first two rows because the entries of the third row  are not yet measured.

The above expressions
 will be used to test globally the unitarity property of the experimental data.

In the following we test our method on  the published data, i.e. we want  to see if our necessary and sufficient criterion, $-1\le\cos\varphi\le 1$, could constrain enough the data. For that we
 use the PDG \cite{Ha} data,  the  fit \cite{HLLL} and its recent up-to-date results \cite{HLLL1} , and consider that a
good starting point for a comparison of the  unitarity triangle approach, almost exclusively used nowadays, with our method is to look at  the central values from the above cited papers. These values are given in relation (\ref{ckm1})
where we used the notation $\sqrt{V}$ to denote the numerical CKM data matrix as it is usually provided. All matrices (\ref{ckm1}) satisfy quite well the stochasticity property (\ref{sto}) as it is seen from the  Table 1.

\begin{eqnarray}\begin{array}{c}
\sqrt{V_{PDG}}=\left(\begin{array}{lll}
0.97485&0.2225&0.00365\\
0.2225&0.974&0.041 \\
0.009&0.0405&0.99915
\end{array}\right)\vspace*{2mm}\\
\sqrt{V_{CKMfG}[5]}=\left(\begin{array}{lll}
0.97504&0.2221&0.0035\\
0.2220&0.97422&0.0408 \\
0.0079&0.04025&0.99917
\end{array}\right)\vspace*{2mm}\label{ckm1}\\
\sqrt{V_{CKMfG}[6]}=\left(\begin{array}{lll}
0.97400&0.2265&0.00387\\
0.2264&0.97317&0.04113 \\
0.00826&0.04047&0.999146\\
\end{array}\right)\end{array}
\end{eqnarray}
Here CKMfG denotes the CKM fitter Group.

 As we said before $\cos\varphi$ depends on the four
 independent parameters we use to obtain it and for comparison we took six groups of independent parameters in the  Table 2.
 It is easily seen that the central values from all  $V_{PDG}$ and $CKMfG$ are not compatible with unitarity, although they are compatible with the double stochasticity property to an acceptable accuracy. More, the data show that  $\cos\varphi$ can take even imaginary values and these values can not be properly processed in the usual approach of the unitarity triangle and  Wolfenstein approximation, which are the starting points of all the present analysis of the CKM data. In  all the  cases it is  $s_{13}$ that causes the trouble.  For the PDG data  $s_{13} =\sqrt{V_{cs}^2+V_{ts}^2- V_{ud}^2}=4.0\times 10^{-3}\, i  $ and, respectively,
$s_{13}=\sqrt{V_{cd}^2+V_{cs}^2+V_{td}^2+V_{ts}^2-1}= 10^{-2}\, i$; for the CKM fitter Group results \cite{HLLL} it has the form $s_{13} =\sqrt{1-V_{ud}^2-V_{us}^2}=5.6\times 10^{-3}\, i $.
 On the other hand such an incompatibility cannot, in principle, be detected by using the Wolfenstein parameterization, because  quantities as  $s_{13}=\sqrt{1-V_{ud}^2-V_{us}^2} $, that may become imaginary in some cases, are usually approximated by  $s_{13}=V_{ub}$ and never appear in the usual approach.
The conclusion is that unitarity requires a very fine tuning between all the entries of the matrix (\ref{pos}) and our method could put strong constraints on the CKM matrix.

\begin{table}
\begin{tabular}{|c|l|l|c|}
\hline
& PDG \cite{Ha}& HLLL \cite{HLLL}&HLLL \cite{HLLL1}\\\hline
$r_1$&-1.48$\times 10^{-4}$&4.37$\times 10^{-5}$&-6.77$\times 10^{-6}$\\\hline
$r_2$&-1.37$\times 10^{-4}$ &5.32$\times 10^{-5}$&8.49$\times 10^{-6}$\\\hline
$r_3$&2.2$\times 10^{-5}$ &2.32$\times 10^{-5}$&-1.22$\times 10^{-6}$ \\\hline
$c_1$&-8.02$\times 10^{-5} $&4.94$\times 10^{-5}$&1.19$\times 10^{-6}$\\\hline
$c_2$&-1.78$\times 10^{-4} $&5.31$\times 10^{-5}$&-8.02$\times 10^{-8}$ \\\hline
$c_3$&-4.96$\times 10^{-6}$ &1.76$\times 10^{-5}$&-6.17$\times 10^{-7}$\\\hline
\end{tabular}\caption
{ $r_i\,\, {\rm and}\,\, c_i,\,\, i=1,2,3$, are the 
 values on the left\\ hand side  in Eqs.(\ref{sto})
calculated using the data (\ref{ckm1})}
\end{table}

For the proper fit we  considered six groups of  four independent parameters,
those appearing in the first column of Table 2,  that lead to 17 different expressions for $\cos\varphi$. We considered that these six groups could be equivalent to the six orthogonality relations implied by unitarity. For minimization of the $\chi^2$ we used the {\em FindMinimum} function provided by {\em Mathematica}. Special care was taken for properly treating the cases that lead to  $\cos\varphi$ values outside the physical range.
\begin{table}
\begin{tabular}{|c|c|c|c|c|}
\hline
 Para's& $\cos\varphi$ & PDG \cite{Ha} & HLLL \cite{HLLL}&HLLL \cite{HLLL1} \\
\hline\hline
$V_{ud},V_{ub},$ &$\cos\varphi^{(1)}$&-1.00237&1.30599&0.483707\\
$V_{cd},V_{cb}$  &$\cos\varphi^{(2)}$&1.10268&0.443955&0.362787\\
&$\cos\varphi^{(3)}$&0.232663&0.506068&0.466784\\
&$\cos\varphi^{(4)}$&0.647108&0.596443&0.458158\\\hline
$V_{ud},V_{us},$ &$\cos\varphi^{(5)}$& 0.348363&-0.376647 $i$&0.473185 \\
$V_{cd},V_{cs}$ &$\cos\varphi^{(6)}$&0.690208&0.130873 $i$&0.459106\\
&$\cos\varphi^{(7)}$&-0.407954&-0.92185 $i$&0.472234\\\hline
$V_{cd},V_{cs},$ &&&&\\
$V_{td},V_{ts}$&$\cos\varphi^{(8)}$&0.452312 $i$&0.577013 &0.459978\\\hline
$V_{ud},V_{ub},$ &$\cos\varphi^{(9)}$&1.02868&0.718312&0.354383\\
$V_{cs},V_{tb}$   &$\cos\varphi^{(10)}$&-0.997118&1.29883&0.484069\\
&$\cos\varphi^{(11)}$&0.236042&0.49466&0.467150\\
&$\cos\varphi^{(12)}$&0.573777&0.87161&0.449737\\\hline
$V_{ud},V_{ub},$ &$\cos\varphi^{(13)}$&-0.943474&0.577013 &0.479405\\
$V_{cs},V_{td}$   &$\cos\varphi^{(14)}$&0.27128&1.30443 &0.462446\\
&$\cos\varphi^{(15)}$&-0.176829&0.503588 &0.558023\\\hline
$V_{ud},V_{cd},$ &$\cos\varphi^{(16)}$&1.15643 $i$&1.07607 &0473158\\
$V_{cs},V_{ts}$   &$\cos\varphi^{(17)}$&0.082041 $i$&0.464743 &0.459054\\\hline
\end{tabular}\caption{All the central data (\ref{ckm1}) coming from  \cite{Ha}, \cite{HLLL} and \cite{HLLL1} are not compatible with unitarity, although the last fit \cite{HLLL1} is considerably better than the published one.  The left column contains the six groups of independent parameters used in our analysis.}\end{table}

We started the fit by using the information provided by the first group of four parameters in Table 2 since we wanted a comparison with the unitarity triangle approach that uses the same information.
With no independent parameters we found a $\chi^2=\chi^2_1+\chi^2_2=3.8 \times 10^{-3}$ and we used the $V_{ij}$ determined parameters to test all the seventeen $\cos\varphi^{(i)}$. We found that nine values, for $i=1,\dots,4,8,13,\dots,17$, are around $0.05$, their mean being $<\cos\varphi>=0.0518$. However we found also a few  discrepancies:   $\cos\varphi^{(5)}=\cos\varphi^{(6)}=0.635$
and  $\cos\varphi^{(7)}=-0.02$; even worse  four values for $i=9,\dots,12$ are outside the physical region. We interpret this phenomenon as showing that the 
use of  only one orthogonality constraint leads to 
 non reliable results.

Taking  into account the   $\cos\varphi^{(i)}$ provided by the second group of independent parameters the fit improves, all the   $\cos\varphi^{(i)}$ being inside the physical region, and  the maximum difference between cosines is of the order $ 5.\times 10^{-2}$ and $\varphi$ is around $58^0$, value compatible to that provided by the CKM fitter Group \cite{HLLL1}.
Including now  in $\chi^2$ all the  the
first twelve  $\cos\varphi^{(i)}$ we found that the results considerably improve, and the difference between $\cos\varphi^{(i)}$ is of the order $2.\times 10^{-4}$. The big surprise was that $\varphi$ takes values close to $90^0$ ! 
For comparison, with our method  we obtained by using  the up-to-date $V_{ij}$ parameters provided  by   the CKM fitter Group \cite{HLLL1} that $0.354383\le \cos\varphi^{(i)} \le 0.558023$, values that are listed in the last column of Table 2,  with a mean  value $ <\cos\varphi> =0.460198$  and $\sigma= 0.0436841$ which leads to a CP phase $\varphi=62.0001^0$, the interval of variation being $59.7428^0\le \varphi\le 65.3853^0$, results that almost  coincide with that given in \cite{HLLL1} that are $50^0\leq\varphi\simeq\gamma\le 72^0 $. In our opinion  these results show the limits for the prediction power of the unitarity triangle approach.

If we use only the values  provided by the first group in Table 2, one gets $\varphi=63.7^0\pm 3.^0$. The difference between the two fits comes from their quality. In the following by using our method we found that   $\chi^2$ takes values in the interval $2.\times 10^{-7}\le \chi^2_1\le 1.7\times 10^{-3}$, while the same expression provides $\chi^2_1=2.95$ when using the last  $V_{ij}$ values obtained by  CKM fitter Group.

We included step by step all the constraints implied by the six groups and obtained three different matrices; they were used to obtain another one by using the convexity property of the unistochastic matrices. Other matrices were obtained by relaxing the condition (\ref{chi2}) by taking all combinations with five and respectively four $V_{ij}$ that provided ${6\choose 1}+{6 \choose 2}=21$ new matrices. The set of these  matrices was considered as 25 independent ``experiments'' on which the statistics was done. For $\cos\varphi$ these lead to $25\times 17= 425$ values that gave $\varphi=89.9962^0 \pm 0.0767^0$. The interval for the $\chi^2_1$ values for all the 25 matrices was shown above and the final results are shown in the  Table 3.
In fact the obtained values can be improved. 

One way to do this is to consider the number of all the possible four independent parameters $V_{ij}$ which is ${9\choose 4}= 126$, but not all are independent; we have to exclude all groups of four parameters that contain all entries from a row or a column, whose number is 36, and also all groups that are not independent because of the unitarity, e.g. $(V_{ud}, V_{ub}, V_{cs}, V_{ts})$, whose number is 9. This leads to $126-36-9=81$ groups and the estimated number of $\cos \varphi$ is about 240. The only trouble is that the fitting procedure is very time consuming even on a good work station.

Looking at the final results we see that  $V_{ij}$ values are not  far from those provided by other fits.
 The striking feature concerns the values 
for the angles of the unitarity triangle, $\alpha, \beta\,\,{\rm and}\,\, \gamma$ , angles that are not obtained from the fit, but from parameters provided by the fit. One sees that with a very good approximation the standard unitarity triangle is almost a rectangle one. We have $\varphi > \gamma$ where their difference is about $4.\times 10^{-2}$.  However this leads to the unexpected result, $\sin 2\alpha \simeq\sin 2\beta$, relation that is not satisfied by the experimental data, which in particular are not very clean.
 We obtain that 
$\sin 2\beta= 0.677$, value that is not far from the world average of the BaBar \cite{BaBar} and Belle \cite{Belle} experiments, that is  $\sin 2\beta=0.736\pm 0.049$ \cite{Fl}.
In fact both angles   $\alpha\,\,{\rm and}\,\, \beta$ are at the lower extremity of the typical ranges for them \cite{Fl},
$70^0 \le \alpha\le 130^0,\quad 20^0\le\beta\le 30^0$. We mention that a rectangle unitarity  triangle was found in \cite{HZ}  but with $\alpha =90^0$.

With our values we find  $Im({U_{CKM}^*}_{ts}{U_{CKM}}_{td})=(1.33 \pm 0.03)\times 10^{-4}$ which is compatible  to that used in \cite{BJ} for determination of the CP-violating ratio $\epsilon'/\epsilon$.

In conclusion we can say that the unitarity triangle results are not reliable and that our approach could outperform by far all the other methods used to reconstruct a CKM unitary matrix from experimental data.
\vskip3mm
{\bf Acknowledgments}. We would like to thank Simone Severini and Karol \.{Z}yczkowski for e-mail correspondence concerning criteria of separating unistochastic matrices from double stochastic ones.
\begin{table}
\begin{tabular}{ lc l}
\hline
Quantity &~~~~~~~ & Central value $\pm$ error\\\hline
\\
$V_{ud}$&&0.974868 $\pm$ 0.00002\\
&&\\
$V_{us}$&& 0.222755  $\pm$  0.00003\\
&&\\
$V_{ub}$&&(3.59529 $\pm$  0.0021)$ 10^{-3}$\\
&&\\
$V_{cd}$&&0.222568 $\pm$  0.000021 \\
&&\\
$V_{cs}$&&0.974049  $\pm$  0.00003 \\
&&\\
$V_{cb}$& &$(4.11362  \pm 0.00285)10^{-2}$\\
&&\\
$V_{td}$ & &$(9.80945  \pm 0.00244) 10^{-3}$ \\
&&\\
$V_{ts}$ &&$(4.01109  \pm 0.00264)10^{-2} $\\
&&\\
$V_{td}$& &0.999147  $\pm$ 0.000028\\
&&\\
$\sin \theta_{12}$&& 0.222725  $\pm$ 0.0000455\\
&&\\
$\sin \theta_{13}$&&$ (3.58334  \pm 0.306) 10^{-3}$\\
&&\\
$\sin \theta_{23}$& &$ (4.11351 \pm 0.03325)10^{-2}$\\
&&\\
J&&$(3.20 \pm 0.27)10^{-5}$\\
\\
$\varphi$&& $89.9962^0 \pm 0.0767^0$\\
&&\\
$\alpha$&&$68.7517^0 \pm 0.02^0$\\
&&\\
$\beta$&&$21.2862^0 \pm 0.02^0$\\
\\
$\gamma$&&$89.9591^0\pm 0.03^0$\\
\\
\hline
\end{tabular}
\caption{Fit results  and errors using the standard input from PDG data. The results show that there is a unitary matrix compatible with the data. The values for $\alpha$, $\gamma$  and $\varphi$ strongly disagree with the previous determinations.}
\end{table}

\end{document}